\documentclass[12pt]{article}
\usepackage{amsfonts}
\usepackage{latexsym}
\usepackage{amsmath,amssymb}
\usepackage{verbatim}
\usepackage{setspace}
\usepackage{color}
\usepackage{physics}
\usepackage{tikz}
\usepackage{mathdots}
\usepackage{yhmath}
\usepackage{cancel}
\usepackage{color}
\usepackage{siunitx}
\usepackage{array}
\usepackage{multirow}
\usepackage{amssymb}
\usepackage{gensymb}
\usepackage{tabularx}
\usepackage[normalem]{ulem}
\usepackage{booktabs}
\usetikzlibrary{fadings}
\usetikzlibrary{patterns}
\usetikzlibrary{shadows.blur}
\usetikzlibrary{shapes}
\usepackage{cite}
\usepackage{hyperref}

\usepackage[textheight=9in, textwidth=6.5in, letterpaper]{geometry}
\def\half{{1\over 2}}
\numberwithin{equation}{section}

\def\e{{\epsilon}}
\def\cs{{\cal S}}

 \def\p{\partial}
 \def\bz{{\bar z}}
 
\def\0{{(0)}}
\def\1{{(1)}}
\def\2{{(2)}}
 
\def\co{{\cal O}}

\def\<{\langle }
\def\>{\rangle }
\def\[{\left[}
\def\]{\right]}
\def\bw{{\bar w}}

\newcommand{\bea}{\begin{eqnarray}}
\newcommand{\eea}{\end{eqnarray}}
\newcommand{\be}{\begin{equation}}
\newcommand{\ee}{\end{equation}}
\newcommand{\ba}{\begin{align}}
\newcommand{\ea}{\end{align}}

\renewcommand{\epsilon}{\varepsilon}

\def\sll{$SL(2, \mathbb{R})_L$}
\def\slr{$SL(2, \mathbb{R})_R$}

   \makeatletter
  \let\over=\@@over \let\overwithdelims=\@@overwithdelims
  \let\atop=\@@atop \let\atopwithdelims=\@@atopwithdelims
  \let\above=\@@above \let\abovewithdelims=\@@abovewithdelims
\renewcommand\section{\@startsection {section}{1}{\z@}%
                                   {-3.5ex \@plus -1ex \@minus -.2ex}
                                   {2.3ex \@plus.2ex}%
                                   {\normalfont\large\bfseries}}

\renewcommand\subsection{\@startsection{subsection}{2}{\z@}%
                                     {-3.25ex\@plus -1ex \@minus -.2ex}%
                                     {1.5ex \@plus .2ex}%
                                     {\normalfont\bfseries}}

\def\p{\partial}

\def\G{\Gamma}

\def \bw {{\bar w}}
\def \bz {{\bar z}}
\def\cS{{\cal S}}

\def\RR{\mathbb{K}}
\def\i{i^\prime}
\def\adz{AdS$_3/\mathbb{Z}$}
\begin{document}
\begin{titlepage}

\unitlength = 1mm
\ \\
\vskip 2cm
\begin{center}
{\LARGE{\textsc{ $(2,2)$ Scattering and the Celestial Torus}}}

\vspace{0.8cm}
Alexander Atanasov$^1$, Adam Ball$^1$, Walker Melton$^1$, Ana-Maria Raclariu$^2$ \\
and Andrew Strominger$^1$
\vspace{1cm}

$^1${\it  Center for the Fundamental Laws of Nature, Harvard University,\\
Cambridge, MA 02138, USA}\\
$^2${\it Perimeter Institute for Theoretical Physics, Waterloo, ON, Canada}

\begin{abstract}
Analytic continuation from Minkowski space to $(2,2)$ split signature spacetime has proven to be a powerful tool for the 
study of scattering amplitudes. Here we show that, under this continuation, null infinity becomes  the product of a null interval with a celestial torus (replacing the celestial sphere) and has only one connected component. Spacelike and timelike
infinity are time-periodic quotients of AdS$_3$. These three components of infinity combine to an $S^3$ represented as a toric fibration  over the interval. Privileged scattering states of scalars organize into \sll $\times$\slr\ conformal primary wave functions and their descendants with real integral or half-integral conformal weights,  giving the normally continuous scattering problem a discrete character.  
 
 \end{abstract}
\vspace{0.5cm}
\end{center}
\vspace{0.8cm}

\end{titlepage}


\tableofcontents
\section{Introduction}
   Scattering amplitudes  in quantum field theory are often defined by analytic continuation 
from $(4,0)$ Euclidean signature to $(3,1)$ Lorentzian signature. This provides an efficient prescription for  the Feynman-diagram singularities encountered in perturbation theory. Moreover, positivity properties in Euclidean space enable powerful non-perturbative instanton and axiomatic analyses.  Euclidean methods have also proven effective in quantum gravity. 
 
 In recent years, however, analytic continuation from Minkowski space to a split $(2,2)$ signature spacetime --- which we shall refer to as Klein space\footnote{After the mathematician Felix Klein, who pioneered the study of these spaces  in the Erlangen Program. } $\RR^{2,2}$ --- has emerged as a complementary and surprisingly effective tool in quantum field theory. An awkward feature of  Euclidean space is that 
 particles cannot  be on-shell. Amplitudes are therefore represented as analytic continuations of sums of off-shell processes, which can both  become inordinately complicated and obscure the underlying physics. Dramatic simplifications have been found in some on-shell descriptions in Klein space \cite{Witten:2003nn,Britto:2004ap,ArkaniHamed:2008yf,Cheung:2008dn,Mason:2009sa,ArkaniHamed:2009si,Britto:2010xq,Elvang:2013cua}. The group-theoretic reduction of the 4D $(2,2)$ Lorentz group to the product of two 1D  conformal groups,  the associated reality of the self duality condition \cite{Ooguri:1991fp},  and the non-degeneracy of massless three-point scattering  also lead to significant  simplifications.

 In quantum gravity in asymptotically flat spacetimes, there are yet further reasons to consider Klein space.  The paucity of generally covariant bulk observables --- and more generally the holographic principle --- suggests that any theory of quantum gravity should be defined by boundary observables. In Euclidean space, the conformal boundary is just a point. It seems  challenging  to formulate  a holographic dual which encodes the richness of 
 asymptotically flat  quantum gravity by observables  at a zero-dimensional point. 
 Here we  find that, in contrast,  Klein space has a rich conformal boundary at infinity, providing a suitable potential home for a holographic dual.   
 
  In section \ref{sec:nsti} we show that the conformal boundary at null infinity in Klein space, denoted $\cal I$,  is the product of a null interval with the Lorentzian signature celestial torus. Both spatial and timelike  infinity $i^0$ and  $i'$ are the product of a disk with a circle and are endowed with the conformal metric of AdS$_3/\mathbb{Z}$. Here the $\mathbb{Z}$-quotient  makes the familiar AdS$_3$ cylinder periodic.  The gluing of the toroidal boundaries of these AdS$_3/\mathbb{Z}$ geometries to the celestial tori at the two ends of $\cal I$ trivializes different cycles of the latter,  giving a toric representation of the ${\cal I}\cup i^0\cup i'$ infinity as $S^3$.  Since $\cal I$ has only one connected component, observables are given by an $\cs$-vector rather than an $\cs$-matrix. 
 The fact that the continuation from Minkowski to Klein space leads to the replacement of the sphere with a torus will perhaps prove  useful for sharpening the concept of a celestial conformal field theory. 
  
Section \ref{sec:sym} reviews the \sll$\times$\slr\  symmetry of Klein space. Expressions are given for $L_n$, $\bar L_n$, $n=-1,0,1$ in a natural basis where $L_0\pm \bar L_0$ generate the compact space and time directions of the celestial torus, as well as for the finite group action on the celestial torus.  The group action preserves the AdS$_3/\mathbb{Z}$ hypersurfaces which are a fixed distance from the origin and foliate Klein space. 

Section \ref{sec:pss} considers conformal basis  wave functions for massless scalars. Single-valuedness on the celestial torus requires that the $L_0$ and $\bar L_0$ eigenvalues are either both integer or both half-integer.   ``$L$-primary" solutions are found  corresponding to highest-weight states annihilated by 
$L_1$ and  $\bar L_1$.  More general solutions are then obtained by taking descendants.  Convolutions of these wave functions with the bulk field operator create states which have an interpretation as $L_0, \bar L_0$ eigenstates of the 1+1D celestial CFT living on a spatial circle of the celestial torus. The fact that the time direction of the torus is periodic is not a problem because $L_0+\bar L_0$ is quantized. We also find lowest-weight solutions annihilated by $L_{-1}$ and  $\bar L_{-1}$, as well as mixed solutions annihilated by $L_{\pm 1}, \bar L_{\mp1}$. 

A striking feature of this construction is that the solutions are labelled by three integers: namely the conformal weights and the levels of the left and right descendants, giving $L$-primary scattering a discrete character. This contrasts with dynamics  on the celestial sphere in Minkowski space, where the conformal basis solutions are labelled by three continuous parameters: a position on the sphere and a continuous complex conformal dimension. The discrete character of celestial scattering in Klein space resonates with several other recent developments. 
Spacetime translations shift conformal weights by a half-integer \cite{Stieberger:2018onx}, so the set of all $L$-primaries and their descendants associated to a given spacetime field form a representation of the Poincar\'e group.\footnote{Unlike the continuous complex highest weight representations discussed in \cite{Pasterski:2016qvg,Pasterski:2017kqt} which are restricted to have the real part of the conformal weight equal to unity and cannot be put in representations of translations. } In gauge theory and gravity, the infinite hierarchy of soft currents appears at negative integer weight, while the positive integer weights appear  related to Goldstone bosons \cite{Pate:2019lpp,Banerjee:2020kaa,Banerjee:2020vnt, ghps}.  Poles  at negative even integer conformal weights in celestial scattering amplitudes were recently shown \cite{Arkani-Hamed:2020gyp} to encode the coefficents in the Wilsonian effective action. These poles characterize much or all of the theory and may be naturally probed by scattering $L$-primaries.

 In section \ref{sec:pso} we construct, as Mellin transforms of plane waves,  modes corresponding  to particles which emerge at a fixed point on the celestial torus. These 
 correspond to ``$H$-primary" operators which are primary with respect to elements $H_1, \bar H_1$ leaving  fixed the point at which the particles emerge. Scattering of such particles takes the form of a correlation function on the celestial torus. 
 We show that $L$-primary wave functions can be expressed as weighted integrals over the torus of $H$-primary wave functions with quantized weights. This is a version of the celestial state-operator correspondence. Hence $L$-primary scattering amplitudes are weighted celestial integrals of Mellin transforms of plane wave scattering amplitudes. We close with a few comments in section \ref{sec:cos}. 
 
\section{Null, spacelike and timelike infinity}
\label{sec:nsti}

In this section we conformally compactify $\RR^{2,2}$ and derive the conformal geometry of null infinity $\cal I$, spatial infinity 
$i^0$ and timelike infinity $\i$. 
The flat metric on $\RR^{2,2}$ is 
\be ds_4^2=dzd\bar z-dwd\bw .\ee
In polar coordinates $z=re^{i\phi}$ and $w=qe^{i\psi}$, this becomes 
\begin{equation}
	ds^2 = -dq^2 - q^2 d\psi^2 + dr^2 + r^2 d\phi^2.
\end{equation}
Now define $q-r = \tan U$, $q+r = \tan V$, giving
\begin{equation}
\label{c2m}
	ds^2 = \frac{1}{\cos^2 U \cos^2 V} \left( -dU dV - \frac{1}{4} \sin^2(V+U) d\psi^2 + \frac{1}{4} \sin^2(V-U) d\phi^2 \right).
\end{equation}
The coordinate ranges are the solid triangle  $-\frac{\pi}{2} < U < \frac{\pi}{2}$ and $|U| < V < \frac{\pi}{2}$, as depicted in figure 1.
Null infinity $\cal I$  is at $V = \pi/2$  where the factor out front blows up. Spacelike (timelike) infinity $i^0$ ($\i$) is the boundary at $U=-{\pi \over 2}$ ($U={\pi \over 2}$). 

Note that, unlike  the case of ${\mathbb M}^{3,1}$, null infinity has only one connected component. This means we cannot define an $\cS$-matrix. Instead we have only an $\cS$-vector in the sense of \cite{Witten:2001kn}. 
It is an amplitude for a collection of incoming particles on $\cal I$ to scatter into nothing --- which they must as there is nowhere to go!  This $\cS$-vector together with a suitable analytic continuation procedure can in principle be used to define an  $\cS$-matrix in ${\mathbb M}^{3,1}$.

\begin{figure}\label{fig:1}
	\tikzset{every picture/.style={line width=0.75pt}} 

	\begin{center}
	\begin{tikzpicture}[x=0.75pt,y=0.75pt,yscale=-1,xscale=1]

	\draw  [fill={rgb, 255:red, 200; green, 200; blue, 200 }  ,fill opacity=0.5 ] (179.5,41.11) -- (399.27,260.76) -- (179.5,260.76) -- cycle ;
	\draw [color={rgb, 255:red, 208; green, 2; blue, 27 }  ,draw opacity=1 ]   (132.61,213.14) -- (179.5,260.76) ;
	\draw [shift={(130.5,211)}, rotate = 45.45] [fill={rgb, 255:red, 208; green, 2; blue, 27 }  ,fill opacity=1 ][line width=0.08]  [draw opacity=0] (10.72,-5.15) -- (0,0) -- (10.72,5.15) -- (7.12,0) -- cycle    ;
	\draw [color={rgb, 255:red, 74; green, 144; blue, 226 }  ,draw opacity=1 ]   (289.89,150.94) -- (225,215.5) -- (179.5,260.76) ;
	\draw [color={rgb, 255:red, 208; green, 2; blue, 27 }  ,draw opacity=1 ]   (228.38,212.12) -- (179.5,260.76) ;
	\draw [shift={(230.5,210)}, rotate = 135.13] [fill={rgb, 255:red, 208; green, 2; blue, 27 }  ,fill opacity=1 ][line width=0.08]  [draw opacity=0] (10.72,-5.15) -- (0,0) -- (10.72,5.15) -- (7.12,0) -- cycle    ;
	\draw [color={rgb, 255:red, 74; green, 144; blue, 226 }  ,draw opacity=1 ]   (272.5,260) .. controls (272.5,232) and (282,172) .. (289.39,150.94) ;
	\draw [color={rgb, 255:red, 74; green, 144; blue, 226 }  ,draw opacity=1 ]   (330.5,261) .. controls (329.5,220) and (325.5,198) .. (289.39,150.94) ;
	\draw [color={rgb, 255:red, 74; green, 144; blue, 226 }  ,draw opacity=1 ]   (289.89,150.94) .. controls (251,159) and (206.5,167) .. (179.5,167) ;
	\draw [color={rgb, 255:red, 74; green, 144; blue, 226 }  ,draw opacity=1 ]   (230.5,261) .. controls (231.5,221) and (277.5,168) .. (289.39,150.94) ;
	\draw [color={rgb, 255:red, 74; green, 144; blue, 226 }  ,draw opacity=1 ]   (289.39,150.94) .. controls (253.5,122) and (208.5,108) .. (179.5,110) ;
	\draw [color={rgb, 255:red, 74; green, 144; blue, 226 }  ,draw opacity=1 ]   (179.5,211) .. controls (224.5,206) and (270.5,164) .. (289.39,150.94) ;
	\draw [color={rgb, 255:red, 74; green, 144; blue, 226 }  ,draw opacity=1 ]   (289.39,150.94) .. controls (251.5,113) and (207.5,76) .. (179.5,76) ;
	\draw [color={rgb, 255:red, 74; green, 144; blue, 226 }  ,draw opacity=1 ]   (370.5,261) .. controls (371.5,235) and (318.5,186) .. (289.89,150.94) ;

	\draw (408.13,249.2) node [anchor=north west][inner sep=0.75pt]   [align=left] {$\displaystyle i^{0}$};
	\draw (159.64,33.67) node [anchor=north west][inner sep=0.75pt]   [align=left] {$\displaystyle i^{'}$};
	\draw (287.98,124.54) node [anchor=north west][inner sep=0.75pt]   [align=left] {$\displaystyle \mathcal{\mathcal I }$};
	\draw (140.5,238.28) node [anchor=north west][inner sep=0.75pt]    {$\textcolor[rgb]{0.82,0.01,0.11}{U}$};
	\draw (207,238.78) node [anchor=north west][inner sep=0.75pt]    {$\textcolor[rgb]{0.82,0.01,0.11}{V}$};

	\end{tikzpicture}
	\end{center}
	\caption{ Toric Penrose diagram for signature $(2,2)$ Klein space. 45$^{\rm o}$ lines are null as usual. A Lorentzian torus is fibered over every point in the diagram. The spacelike cycle of the torus degenerates along the timelike line $U=V$, while the timelike cycle degenerates along the spacelike line $U=-V$. Neither cycle degenerates at null infinity $\cal I$ which is the interval $- {\pi \over 2}<U< {\pi \over 2},~V={\pi \over 2}$. Spacelike infinity $i^0$ is at $(U,V)=(-{\pi \over 2},{\pi \over 2})$ and has the conformal geometry of signature $(1,2)$ \adz. Timelike infinity $\i$ is at $(U,V)=({\pi \over 2},{\pi \over 2})$ and has the conformal geometry of signature $(2,1)$ \adz. The blue lines are lines of constant $w\bw -z\bz$ with $\tau=0$ at $U=0$. }\end{figure}
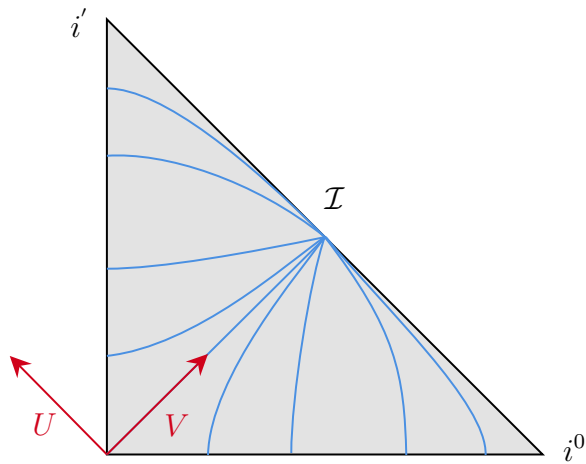

$\cal I$ is parameterized by the null coordinate $-{\pi \over 2}<U<{\pi \over 2}$ and the periodic coordinates $\psi$ and $\phi$. Taking $V\to {\pi \over 2}$ while rescaling \eqref{c2m} by $\cos^2 V$ one finds the conformal metric on $\cal I $ to be the square, Lorentzian torus
\begin{equation}
		ds^2_{\mathcal I} = -d\psi^2 + d\phi^2, \quad \psi\sim \psi+2\pi,~ \phi\sim \phi+2\pi.
\end{equation}
Hence $\cal I$ is the product of the  celestial torus  with a null interval. 

Now we turn to $i^0,\i$. Since the boundary of a boundary is nothing, we must be able to glue these to $\cal I$ to get an $S^3$ which is the topological boundary of  $\RR^{2,2}$.  $S^3$ is topologically represented in toric geometry as a torus fibration over the interval in which one of the two torus cycles shrinks to zero at one end of the interval, and the other at the other end. Then there are no non-contractible cycles. In order to complete $\cal I$ to $S^3$ in this manner, the $i^0,~\i$ ``caps''
must both be topologically the product of a disk and a circle. We now show that this is indeed the case and moreover that the conformal geometry on each cap is \adz.

Following procedures which are standard for $\mathbb{M}^{3,1}$ \cite{Ashtekar:1978zz,Ashtekar:1991vb, deBoer:2003vf,Campiglia:2015qka,Henneaux:2018cst}, we resolve $\i$ by taking the $\tau\to \infty$ limit of	
the  signature $(2,1)$ surface 
\be z\bz-w\bw=-\tau^2.\ee
We denote the two regions of  $\RR^{2,2}$ with positive or negative $z\bz-w\bw$ by $\RR^{2,2\pm}$. Coordinates covering the region $\RR^{2,2-}$, which contains $\i$, are 
\bea z&=&\tau e^{i\phi} \sinh \rho ,\cr
w&=&\tau e^{i \psi} \cosh \rho .
\eea
The inverse relations are 
\be
\begin{split}
\tau &= \sqrt{w\bw - z\bz}, \qquad \tanh \rho = \sqrt{\frac{z\bz}{w\bw}}, \\
e^{i\phi} &= \sqrt{z/\bz},  \qquad\qquad\quad~~ e^{i\psi} = \sqrt{w/\bw}.
\end{split}
\ee
The Klein space metric in these coordinates is 
\be ds_4^2=-d\tau^2+\tau^2ds_3^2,\ee
where
\be ds_3^2=-\cosh^2 \rho \, d\psi^2+\sinh^2 \rho \, d\phi^2 +d\rho^2 \ee
is the conformal geometry of $\i$. We recognize it as the standard metric on \adz, where the $\mathbb Z$ acts as the time-like quotient 
$\psi\to \psi+2\pi$.

A similar construction for $i^0$ begins with $(\tilde \tau, \tilde \rho, \phi, \psi)$ covering  $\RR^{2,2+}$ with 
$ z\bz-w\bw=+\tilde \tau^2$:
\bea z&=&\tilde \tau e^{i\phi} \cosh\tilde \rho ,\cr
w&=&\tilde \tau e^{i \psi} \sinh \tilde \rho .
\eea
The inverse relations are
\be 
\begin{split}
\tilde{\tau} &= \sqrt{z\bz - w\bw}, \qquad \tanh \tilde{\rho} = \sqrt{\frac{w\bw}{z\bz}}, \\
e^{i\phi} &= \sqrt{z/\bz}, \qquad\qquad\quad~~ e^{i\psi} = \sqrt{w/\bw}.
\end{split}
\ee
One finds 
\be ds_4^2=d\tilde\tau^2-\tilde \tau^2ds_3^2,\ee
\be ds_3^2=-\cosh^2 \tilde \rho \, d\phi^2+\sinh^2 \tilde \rho \, d\psi^2 +d\tilde \rho^2. \ee
We see that the non-contractible loop in the \adz\  factor is now $\phi$ instead of $\psi$ and spacelike instead of timelike in the $\RR^{2,2\pm}$ embedding space. Hence gluing the conformal geometries of the two \adz\  caps to $\cal I$ trivializes both cycles of the celestial torus
and the  full topology  of infinity is $S^3$.

\section{Symmetries}
\label{sec:sym}

The ``Lorentz  group" of $\RR^{2,2}$ is $SO(2,2) \cong \frac{SL(2,\mathbb{R})_L \times SL(2,\mathbb{R})_R}{\mathbb{Z}_2}$, where the $\mathbb{Z}_2$ is generated by $-1_L \times -1_R$. 
The spin group is the double cover $SL(2,\mathbb{R})_L \times SL(2,\mathbb{R})_R$. 
The symmetry is generated on Klein space by (real combinations of) the six Killing vector fields
\begin{equation}
\label{gens}
\begin{split}
L_1 & = \bz \p_w + \bw \p_z, \qquad\qquad\qquad\qquad~~ \bar{L}_{1} = z\p_w + \bw\p_\bz, \\
L_0 &= \half \left( z\p_z + w\p_w - \bz\p_\bz - \bw\p_\bw \right),\quad \bar{L}_0 = \half \left(- z\p_z + w\p_w + \bz\p_\bz - \bw\p_\bw \right), \\
L_{-1} &= -z\p_\bw - w\p_\bz, \qquad\qquad\qquad\quad~
\bar{L}_{-1} = -\bz\p_\bw - w\p_z.
\end{split}
\end{equation}
In $\RR^{2,2-}$ we may also write 
\begin{equation}
\label{gns}
\begin{split}
L_1 &= \half e^{-i\psi-i\phi} \left( \partial_{\rho} - i\tanh \rho \, \p_\psi - i\coth \rho \, \p_\phi \right), \\
L_0 &= -{i \over 2} \left(\partial_{\psi} + \partial_\phi \right), \\
L_{-1} &= \half e^{i\psi + i\phi} \left( -\p_{\rho} - i\tanh \rho \, \p_\psi - i\coth \rho \, \p_\phi \right), \\
\bar{L}_{1} &= \half e^{-i\psi + i\phi} \left( \p_{\rho} - i\tanh \rho \, \p_\psi + i\coth \rho \, \p_\phi \right), \\
\bar{L}_0 &= -{i \over 2} \left( \partial_{\psi} - \partial_\phi \right), \\
\bar{L}_{-1} &= \half e^{i\psi-i\phi} \left( -\p_{\rho} - i\tanh \rho \, \p_\psi + i\coth \rho \, \p_\phi \right),
\end{split}
\end{equation}
while on  $\RR^{2,2+}$
\begin{equation}
\label{gns1}
\begin{split}
L_1 &= \half e^{-i\psi - i\phi} \left( \p_{\tilde \rho} - i\coth \tilde{\rho} \, \p_\psi - i\tanh \tilde{\rho} \, \p_\phi \right), \\
L_0 &= -{i \over 2} \left( \p_\psi + \p_\phi \right), \\
L_{-1} &= \half e^{i\psi + i\phi} \left( -\p_{\tilde \rho} - i\coth {\tilde \rho} \, \p_\psi - i\tanh {\tilde \rho} \, \p_\phi \right), \\
\bar{L}_{1} &= \half e^{-i\psi + i\phi} \left( \p_{\tilde \rho} - i\coth {\tilde \rho} \, \p_\psi + i\tanh {\tilde \rho} \, \p_\phi \right), \\
\bar{L}_0 &= -{i \over 2} \left( \p_\psi - \p_\phi \right), \\
\bar{L}_{-1} &= \half e^{i\psi - i\phi} \left( -\p_{\tilde \rho} - i\coth {\tilde \rho} \, \p_\psi + i\tanh {\tilde \rho} \, \p_\phi \right).
\end{split}
\end{equation}
In either case on the boundary at $\rho \to \infty$ (or $\tilde{\rho} \to \infty$) these reduce to the familiar circle action
\begin{equation}
\label{gns1}
L_n = -{i \over 2}  e^{-in(\psi +\phi)} \left(  \p_\psi +  \p_\phi \right), \quad \bar{L}_{n} = -{i \over 2} e^{-in(\psi -\phi)} \left( \p_\psi - \p_\phi \right),\end{equation}
for $n=-1,0,1$. The $L_n$ obey (for all $\rho$ or $\tilde{\rho}$) the $SL(2, \mathbb{R})$ Lie bracket algebra 
\begin{equation}
\label{cr}
\[L_n, L_{m}\] = (n-m)L_{m+n}, 
\end{equation}
and similarly for the $\bar L_n$.

\adz\ is the $SL(2, \mathbb{R})$ group manifold which admits an \sll$\times$\slr\  group action.  The generators above leave fixed the \adz\  hypersurfaces of constant $w\bw-z\bz$. $L_0-\bar L_0$ generates AdS rotations and $L_0+\bar L_0$ generates AdS global time translations for $w\bw-z\bz=\tau^2$, while for $z\bz-w\bw=\tilde{\tau}^2$ it is the other way around. In either  case because of the mod $\mathbb Z$ quotient, the eigenvalues of $L_0\pm\bar L_0$ are separately quantized. This is standard in 
 $SL(2, \mathbb{R})$ representation theory, but differs from familiar string theory applications in which one works on the simply-connected universal cover AdS$_3$  of \adz, and only $L_0-\bar L_0$ is quantized.

$SO(2,2)$ acts faithfully on the celestial torus. We define the null angles
\be
x^\pm \equiv \psi \pm \phi.
\ee
While $\psi, \phi$ naturally parametrize the cycles of the celestial torus, the symmetry group acts more simply on $x^\pm$. In particular \sll \, acts only on $x^+$, while \slr \, acts only on $x^-$. The price for working with $x^\pm$ is that their periodicity properties are not independent. Rather one has
\be (x^+, x^-) \sim (x^+ + 2\pi, x^- + 2\pi) \sim (x^+ + 2\pi, x^- - 2\pi). \ee
Finite elements of \sll \, act as M\"obius transformations on $\tan \frac{x^+}{2}$ by
sending $x^+ \to x^{+'}$ such that
\be \tan \frac{x^{+'}}{2} = \frac{a \tan \frac{x^+}{2} + b}{c \tan \frac{x^+}{2} + d} \ee
with $ad-bc=1$. 
Note that $\tan \frac{x^+}{2} = \tan \frac{x^+ + 2\pi}{2}$ despite the fact that $(x^+, x^-)$ and $(x^+ + 2\pi, x^-)$ are distinct points, so  $\tan \frac{x^+}{2}$ is not a good coordinate on the whole torus. 

\section{Primary scalar states}
\label{sec:pss}

In this section we construct the highest- and lowest-weight conformal primary wave functions for a massless scalar.

Solving the massless scalar wave equation $\square \Phi=0$ in a conformal basis reduces to finding representations of \sll$\times$\slr\  as functions on the $SL(2, \mathbb{R})$ group manifold \adz. On $\mathbb{K}^{2,2-}$ the equation separates as $\Phi(\tau,\psi,\phi,\rho)=\Phi_1(\tau)\Phi_3(\psi,\phi,\rho)$ and can be written  
\bea \label{sone}
	\frac{1}{\tau}(\partial_{\tau} \tau^3 \partial_{\tau})\Phi_1 &=&K\Phi_1, \\
	\label{stwo}
        \nabla^2_3 \Phi_3&=&K\Phi_3,
\eea
where the separation constant $K$ can be anything at this point. The first equation
\eqref{sone}
has two power law solutions which depend on $K$. 
\eqref{stwo} is the wave equation for a scalar of mass $m^2=K$ on \adz.  In a standard basis, $L_0+\bar L_0$ generates time translations, while $L_0-\bar L_0$ generates space rotations.  Both must be integers, implying that $L_0$ and $\bar L_0$ are either both integers or both half-integers.\footnote{ In the  familiar case of $AdS_3$, $\psi$ is not periodically identified and $L_0+\bar L_0$ is not quantized. }
\eqref{stwo} can be rewritten in terms of either the \sll\ or \slr\ Casimirs on \adz
\be (4\bar L_0^2 - 2\bar L_{-1}\bar L_1 - 2\bar L_{1}\bar L_{-1})\Phi_3=(4L_0^2 - 2L_{-1}L_1 - 2L_{1} L_{-1})\Phi_3=K\Phi_3. \ee
Here we consider conformal primary solutions in a basis of $(L_0, \bar L_0)$ eigenstates.  These obey the eigenvalue condition 
\begin{equation}\label{eg}
L_0 \Phi_3 = h \Phi_3, \quad \bar{L}_0 \Phi_3 = \bar{h} \Phi_3
\end{equation}
for integer or half-integer $(h,\bar h)$, as well as the highest-weight condition 
\be\label{hw}  L_1\Phi_3=\bar L_1 \Phi_3=0.\ee
We refer to these as ``$L$-primary'', to distinguish them from operator-type primaries discussed in the next section. 
Commuting $L_1$ and $\bar L_1$ to the right where they annihilate $\Phi_3$, the wave equation reduces to \eqref{hw}
together with
\be K = 4h(h-1) = 4\bar h (\bar h-1) \ee
for some integer or half-integer $(h, \bar h)$ eigenvalues of $L_0,\bar L_0$. \eqref{sone} then has two solutions
\be \Phi_1 = \tau^{-2h},~~~~ \widetilde \Phi_1= \tau^{2h-2},\ee
which are indirectly related by the shadow transform. 
Moreover the highest-weight conditions \eqref{hw} imply
\be h=\bar h \ee
together with 
\be \p_{\rho} \Phi_3 + 2h \tanh \rho \, \Phi_3=0.\ee
This is solved by
\be
\Phi_3 \propto {e^{2ih\psi} \over \cosh^{2h}\rho}.
\ee
Putting this together we have a pair of conformal primary solutions in $\RR^{2,2-}$ for every half-integer value of $h$,
\be\label{Lp} \Phi^{++}_h={e^{2ih\psi} \over \cosh^{2h}\rho}\tau^{-2h},  ~~~~~~~   \widetilde \Phi^{++}_h={e^{2ih\psi} \over \cosh^{2h}\rho}\tau^{2h-2} .\ee
We can construct descendant  solutions by acting with $L_{-1}, \bar{L}_{-1}$ on $\Phi^{++}_h$. However we still have to match this to a solution on $\RR^{2,2+}$. For this purpose it is easiest  to work in terms of the $(z,\bz,w,\bw)$ coordinates. Then we find 
\be\label{ppp}  \Phi^{++}_h = \bw^{-2h}, \quad \widetilde \Phi^{++}_h = \bw^{-2h} (w\bw - z\bz)^{2h-1}.\ee
 The equation of motion
\be \square \Phi = 4(\p_z\p_\bz -\p_w\p_\bw) \Phi = 0 \ee
has   $\p_{\bar{w}}^{2h}\delta^{(2)}(w)$ sources at  $w=0$ for positive $h$, which may be important or need  regulation in some applications.  The singularity along the  light cone of the origin $z\bz=w\bw$ can be regulated with a $\pm i\epsilon$ prescription,  a choice of which may be necessary for example to define scattering amplitudes. Similar  regulators  are likely  needed in solutions below but will not be analyzed herein.   Near  $\cal I$ at $V={\pi \over 2}$ one finds 
\be 
\Phi^{++}_h \to (\pi-2V)^{2h}e^{2ih\psi},\quad \widetilde \Phi^{++}_h \to  (\pi-2V)e^{2ih\psi}(2\tan U)^{2h - 1}.
\ee

We could also consider lowest-weight solutions obeying 
\be\label{prm}  L_{-1} \Phi = \bar L_{-1} \Phi = 0. \ee
Inspection of \eqref{gens} immediately reveals that the complex conjugates 
\be \Phi^{--}_h =( \Phi^{++}_{h})^* = w^{-2h}, \quad \widetilde \Phi^{--}_h=(\widetilde \Phi^{++}_{h})^* = w^{-2h} (w\bw - z\bz)^{2h-1},  \ee
obey \eqref{prm} and 
\be L_0 \Phi^{--}_h=\bar L_0  \Phi^{--}_h =-h \Phi^{--}_h ,\quad  L_0 \widetilde\Phi^{--}_h=\bar L_0 \widetilde \Phi^{--}_h =-h\widetilde \Phi^{--}_h.\ee

There are further  mixed solutions obeying
\be\label{gkl} L_{1}\Phi = \bar L_{-1}\Phi = 0. \ee
Again from \eqref{gens} we see that under the exchange $z\leftrightarrow w$, we have
\be L_n\leftrightarrow L_n, \quad \bar{L}_n \leftrightarrow -\bar{L}_{-n}. \ee
It follows that
\be \label{pm} \Phi^{+-}_h  = \bz^{-2h}, \quad \widetilde \Phi^{+-}_h=\bz^{-2h} (w\bw - z\bz)^{2h-1}, \ee
obey \eqref{gkl} and 
\be L_0 \Phi^{+-}_h=- \bar L_0  \Phi^{+-}_h =h \Phi^{+-}_h ,\quad  L_0 \widetilde\Phi^{+-}_h=- \bar L_0 \widetilde \Phi^{+-}_h = h \widetilde \Phi^{+-}_h.\ee

Finally the other class of mixed solutions
\be\label{gkl2} L_{-1}\Phi=\bar L_1 \Phi = 0 \ee 
is given by
\be\label{mp}\Phi^{-+}_h  = z^{-2h}, \quad \widetilde \Phi^{-+}_h=z^{-2h} (w\bw - z\bz)^{2h-1}.\ee
The full \sll$\times$\slr\  multiplets can for all cases be obtained by suitable actions of $L_{\pm 1}, \bar L_{\pm 1}$.

\section{Primary scalar operators} 
\label{sec:pso}

In the previous section we constructed wave functions whose convolutions with bulk field operators create states in the $(1,1)$ celestial CFT on the Lorentzian torus. These are $L$-primary with respect to the standard \sll$\times$\slr\ action, diagonalizing both time translations and space rotations. 
 
The $(1,1)$ CFT also contains local operators acting at points on the torus 
\be \co_{h,\bar h}(\hat x^+, \hat x^-),~~~~\hat x^\pm=\hat \psi \pm \hat \phi, \ee
which are $H$-primary rather than $L$-primary \cite{Maldacena:1998bw}. They are annihilated by the raising operators in the basis that diagonalizes boosts towards $(\hat x^+,\hat x^-)$. This basis is 
\be
\begin{split}
H^{\hat x}_{0} &= \frac{1}{2}\left( e^{i \hat x^+} L_1 - e^{-i \hat x^+} L_{-1}\right),\quad
H_{\pm 1}^{\hat x }= iL_0 \mp \frac{i}{2}\left(  e^{i\hat  x^+} L_1 + e^{-i \hat x^+} L_{-1}\right) ,\\
\bar{H}^{\hat x}_{0} &=\frac{1}{2}\left( e^{i \hat x^-} \bar{L}_1 - e^{-i \hat x^-} \bar{L}_{-1}\right),\quad
\bar{H}_{\pm 1}^{\hat x} = i\bar{L}_0 \mp \frac{i}{2}\left(  e^{i \hat x^-} \bar{L}_1 + e^{-i \hat x^-} \bar{L}_{-1}\right).
\end{split}
\ee
These obey the commutation relations
\be 
\begin{split}
	[H_n, H_m] = (n - m) H_{n + m}, ~~[\bar{H}_n, \bar{H}_m] = (n - m) \bar{H}_{n + m}.
\end{split}
\ee
Analogous primary  operators were constructed in Minkowski space, where they live on the sphere rather than the torus, as Mellin transforms of momentum space field operators in \cite{Pasterski:2016qvg}. The construction is easily continued to Klein space. Let us write in $(z, \bz, w, \bw)$ coordinates
\bea 
\label{pX}
p&=& \omega \hat p(\hat{x})=\omega(e^{i\hat \phi},e^{-i\hat \phi},e^{i\hat \psi},e^{-i\hat \psi}),\\
X&=&(re^{i \phi},re^{-i\phi},qe^{i \psi},qe^{-i \psi}),\eea
so that $p^2=0$ and
\bea \hat p(\hat{x}) \cdot X &=& r\cos(\hat \phi - \phi) - q\cos(\hat \psi - \psi)\cr
&=& (r - q)\cos \frac{\hat x^+ - x^+}{2}\cos \frac{\hat x^- - x^-}{2} + (r + q)\sin \frac{\hat x^+ - x^+}{2}\sin \frac{\hat x^- - x^-}{2},\eea
where $\hat{x} = (\hat{x}^+, \hat{x}^-).$
As usual the Mellin transform gives\footnote{In Minkowski space the  $\pm i\e$ prescription at $\hat p\cdot X=0$ distinguishes ingoing and outgoing solutions. In Klein space changing the sign in front of $\e$ is equivalent to changing the sign of $\hat p$ and so does not give a new solution, in accord with the fact that ${\cal I}$ has only one connected component.
In the case when $2h-1$ is a negative integer --- which is related to soft currents in the spin-one case ---  the wave functions should be normalized so as  to cancel the $\G$-function singularities \cite{Guevara:2019ypd,Pate:2019mfs,Puhm:2019zbl, Adamo:2019ipt}. 
}
\be\label{ml} \varphi_{h}(X; \hat{x})=\int_0^\infty d\omega \omega^{2h-1}e^{ i\omega\hat p\cdot X}={e^{- \pi i h}\G(2h)\over (\hat p\cdot X )^{2h}}.\ee
These obey, by construction, the wave equation as well as
\bea\label{edf}  H^{\hat x}_{1}\varphi_{h}(X; \hat x) &=& \bar H^{\hat x}_{1}\varphi_{h}(X; \hat x)=0,\\
H^{\hat x}_{0}\varphi_{h}(X; \hat x) &=& \bar H^{\hat x}_{0}\varphi_{h}(X;\hat x) = h\varphi_{h}(X;\hat x),\\
H^{\hat x}_{-1}\varphi_{h}(X;\hat x) &=& -2\hat \p_+\varphi_{h}(X;\hat x),\\
\bar H^{\hat x}_{-1}\varphi_{h}(X;\hat x) &=& -2\hat \p_-\varphi_{h}(X;\hat x).\eea
Scattering amplitudes of particles with these wave functions are  Mellin transforms of plane wave amplitudes, and are identified with conformal primary correlation functions on the celestial torus. 
These wave functions have branch cuts for generic $h$ and are periodic in both the time and space directions $\psi$ and $\phi$. One may also consider the shadows of these solutions
\be  \widetilde \varphi_{h}(X;\hat x) =\varphi_{h}(X;\hat x)(X^2)^{2h-1},\ee
which obey \eqref{edf}.
 
We can put our $(1,1)$ CFT on the Lorentzian cylinder just as well as the Lorentzian torus, and it is instructive to see how they are related. In a conventional $(1,1)$ CFT on the cylinder there is a canonical map from primary operators at a point to operator modes on the circle, given by an integral over a causal diamond
\be 
\label{me}
O_{m,n} =\int_0^{2\pi} d\hat{x}^+\int_0^{2\pi}d\hat{x}^-e^{-im\hat{x}^+ - in\hat{x}^-}\mathcal{O}_{h, \bar{h}}(\hat{x}^+,\hat{x}^-),
\ee
where $h-\bar{h}\in\mathbb{Z}$. Spatial periodicity requires $m-n\in\mathbb{Z}$. In order that our modes create the primary and descendant states associated with $\mathcal{O}_{h, \bar{h}}$, we need $m, n \in \mathbb Z - h$. When we instead consider this mode expansion on the torus, the timelike periodicity further requires $m+n\in\mathbb{Z}$, which means that the only consistent operators of the type \eqref{me} arise from $H$-primaries with $h + \bar{h} \in \mathbb{Z}$. Accordingly we henceforth restrict to $h\pm\bar h\in \mathbb{Z}$.

The analog of this map (on the torus now) at the level of the wave functions \eqref{ml} is
\be \label{dmd} \Phi_{m,n}(X) = \int_0^{2\pi}d\hat{x}^+\int_0^{2\pi}d\hat{x}^-e^{-im\hat{x}^+ - in\hat{x}^-}\varphi_{h}(X; \hat{x}^+,\hat{x}^-).\ee
Using 
\be L_0=-{i \over 2}(H^{\hat x}_1+H^{\hat x}_{-1}),  \quad L_{\pm 1}={e^{\mp i\hat x^+} \over 2}(iH^{\hat x}_1- iH^{\hat x}_{-1}\pm 2H^{\hat x}_0), \ee
together with \eqref{edf}, and integrating by parts with respect to $\hat x^+$ one finds the standard mode relations
\bea L_1 \Phi_{m,n}&=&(h-1-m)\Phi_{m+1,n},\\
L_0 \Phi_{m,n}&=&-m\Phi_{m,n},\\ L_{-1} \Phi_{m,n}&=&(1-h-m)\Phi_{m-1,n}. \eea
Any solutions $\Phi$ constructed from  linear combinations of $\varphi_{h}(X;\hat x)$ obey
\be (L_0(L_0-1) -  L_{-1} L_1 )\Phi=(L_0(L_0+1) -  L_{1} L_{-1} )\Phi= h(h-1)\Phi. \ee
Hence highest-weight solutions obeying $L_1\Phi=0$ have $m=-h$ or $m=h-1$, while lowest-weight solutions obeying $L_{-1}\Phi=0$ have $m=h$ or $m=1-h$. Similar relations hold for the $\bar L_n$. The highest-weight solution with $m=n=-h$ is 
\bea
\label{HtoL} \Phi_{-h,-h} &=&\int_0^{2\pi}d\hat{x}^+\int_0^{2\pi}d\hat{x}^-e^{ih(\hat{x}^++\hat{x}^-)}\varphi_{h}(X;\hat{x}^+,\hat{x}^-)\\
&=&2^{2 + 2h} \pi^2 e^{i\pi h} \Gamma(2h) \bw^{-2h} \propto \Phi_h^{++}, \eea
in agreement with  \eqref{ppp}. The integral here is performed in appendix \ref{int}.  The poles at negative half-integer $h$ are inherited from the normalization of the wave functions \eqref{ml} and can be absorbed by a redefinition of the wave functions resulting in finite amplitudes \cite{Pate:2019mfs,Puhm:2019zbl, Adamo:2019ipt}.

For $h>0$, taking descendants of the primary generates the standard infinite-dimensional unitary \sll$\times$\slr\ representation. 
For $h<0$, after taking $2h$ descendants (on either left or right) we reach $m=h$ and the representation terminates. This is a non-unitary finite-dimensional representation. 

Similarly for $m=n=h$ we have 
\bea \label{lwLp} \Phi_{h,h} &=&\int_0^{2\pi}d\hat{x}^+\int_0^{2\pi}d\hat{x}^-e^{-ih(\hat{x}^++\hat{x}^-)}\varphi_{h}(X;\hat{x}^+,\hat{x}^-)\\
&=&2^{2 + 2h} \pi^2 e^{i\pi h} \Gamma(2h) w^{-2h} \propto \Phi_h^{--}.  \eea
This is a lowest-weight solution. The representations are filled out by acting with powers of $L_1, ~\bar L_1$. 
Mixed primary solutions can also be obtained by taking $(m,n)$ as $(h,-h)$ or  $(-h,h)$.

\section{Comments on scattering}
\label{sec:cos}
 
Celestial $\cS$-vector elements of particles with Klein space $H$-primary wave functions  are given by  Mellin 
transforms of momentum space $\cS$-vector elements. The $k$th  external particle is labeled by 3 continuous parameters: $h_k, x^+_k,x^-_k$. They take the form of CFT correlation functions  on a Lorentzian torus.
 
  Celestial $\cS$-vector elements of particles with  $L$-primary wave functions, and their descendants,   can also be computed from  momentum space $\cS$-vector elements, with the additional weighted integral over the celestial torus given in \eqref{me}. The $k$th  external particle is labeled by 3 {\it discrete} parameters:  $h_k$ and the left and right levels of the descendant. 
  It is interesting  that the $L$-primary scattering problem has a discrete character.  This resonates with the results of \cite{Arkani-Hamed:2020gyp} 
 where it was shown, for the Minkowskian  four-particle  amplitude,  that the Wilsonian  coefficients are encoded in poles  at discrete integral conformal weights. These  are likely probed by $L$-primary scattering amplitudes.  
 We defer a more detailed analysis of properties of the $L$-primary solutions to future work. 

\section*{Acknowledgments}
We are grateful to Nick Agia, Nima Arkani-Hamed, Alfredo Guevara, Mina Himwich, Noah Miller,  Monica Pate and E. Witten for useful conversations, as well as to Nima Arkani-Hamed for an interesting  history lesson on Felix Klein. 
This work was supported by DOE grant de-sc/0007870,  by Gordon and Betty Moore Foundation and John Templeton Foundation grants via the Black Hole Initiative and by a Hertz Fellowship to A.A. A.R. is supported by the Stephen Hawking Postdoctoral Fellowship at Perimeter Institute. Research at Perimeter Institute is supported in part by the Government of Canada through the Department of Innovation, Science and Industry Canada and by the Province of Ontario through the Ministry of Colleges and Universities.	 	
 
\appendix
\section{Mapping $H$-primaries to $L$-primaries}
\label{int}

In this appendix we evaluate the integral \eqref{HtoL} mapping conformal $H$-primary solutions to $L$-primary ones. We start with
\be 
\label{int}
\begin{split}
I  &=  \int_0^{2\pi} \int_0^{2\pi} d \hat x^+ d \hat x^- e^{-i m \hat{x}^+} e^{-i n \hat{x}^-} \frac{e^{-i\pi h}\Gamma(2h)}{ ( \hat{p} \cdot X)^{2h}}.
\end{split}
\ee
This integral consists of points inside the causal diamond, covering half of the celestial torus. It can be related to an integral over the full torus  by noticing that the integral over the other half is obtained from the integral over the diamond by shifting $\hat{x}^+ \rightarrow \hat{x}^+ + 2\pi$ for fixed $\hat{x}^-.$ This transformation amounts to taking $\hat{p}$ to its antipodal point $ -\hat{p}$.
Under this transformation,
\be
	e^{- i m \hat x^+ - i n \hat x^-} \Phi_h(\hat x^+, \hat x^-) \to e^{-2 \pi i h} e^{- i m \hat{x}^+ - i n \hat{x}^-} e^{\pm 2 \pi i h} \Phi_h(\hat x^+, \hat x^-).
\ee
For $h \in \frac{1}{2} \mathbb{Z}$ the phases on the RHS cancel in which case the integrals over the diamond and its complement are equal. The integral over the causal diamond can then be replaced by half the integral over the torus.
The change of variables to $\hat{\psi}, \hat{\phi}$ then leads to
\be 
I = \int_{T^2} d\hat{\psi} d\hat{\phi} \, e^{-i (m + n) \hat{\psi}} e^{-i (m - n) \hat{\phi}} \frac{e^{-i\pi h}\Gamma(2h)}{\left[r \cos(\hat{\phi} - \phi) - q \cos(\hat{\psi} - \psi)\right]^{2h}}.
\ee
Setting  $m = n$ and 
\be 
 w = e^{i(\psi- \hat{\psi})}, \quad z = e^{i(\hat{\phi} - \phi)},
\ee
we find
\be 
\begin{split}
I = -2^{2h} e^{-2 im \psi}\oint \frac{dw}{w} \oint \frac{dz}{z} w^{2m} \frac{e^{-i\pi h}\Gamma(2h)}{\left[r (z + z^{-1})- q(w + w^{-1})\right]^{2h}},
\end{split}
\ee
where the contours are the unit circles $|z| = |w| = 1.$
For fixed $w$ and $|r(z + z^{-1})| < |q(w + w^{-1})|$ we can expand the denominator \footnote{It can be shown that the analogous expansion for $|r(z + z^{-1})| > |q(w + w^{-1})|$ gives the same result for $m = -h$. }
\be 
I =  -2^{2h} e^{-2 im \psi} e^{-i\pi h}\Gamma(2h) \oint \frac{dw}{w} \oint \frac{dz}{z} w^{2m} \sum_{k =0}^{\infty}(- q)^{-2h}(w + w^{-1})^{-2h}  \left(-\frac{r}{q}\frac{z + z^{-1}}{w + w^{-1}} \right)^k \left( \begin{matrix}
-2h\\
k
\end{matrix}\right).
\ee
 and first evaluate the $z$ integral, 
\be 
\oint \frac{d z}{z} (z + z^{-1})^k = 2\pi i \begin{cases} \frac{\Gamma(k + 1)}{\Gamma(k/2 + 1)^2}, ~~ &k \in 2\mathbb{Z}_+,\\
0,~~ &{\rm else}.
\end{cases}
\ee
Then, upon a redefinition of the summation variable,
\be 
\label{Int}
\begin{split}
I &=  -2^{2h} e^{-2 im \psi} e^{-i\pi h}\Gamma(2h) 2\pi i\sum_{k = 0}^{\infty} \oint \frac{dw}{w} w^{2m} (w + w^{-1})^{-2h -2k}(-q)^{-2h} \left(-\frac{r}{q}\right)^{2k}  \\
&\times \frac{\Gamma(-2h + 1)}{\Gamma(1 + k)^2\Gamma(-2h - 2k + 1)}. 
\end{split}
\ee
We now easily see that the remaining integral above simplifies and gives us the solutions in \ref{ppp}. First set $m = -h$. The integral over $w$ reduces to
\be 
\oint \frac{dw}{w} (w^2 + 1)^{-2h-2k} w^{2k}.
\ee
For $k > 0$, there are no poles inside the contour $|z| = 1$, so this integral vanishes.\footnote{The branch cuts at $\pm i$ can be pushed away from the contour by an $i\epsilon$ prescription.} The only term that survives in \eqref{Int} is the $k = 0$ one which gives
\be 
\label{-h}
I_{m = -h} = 4\pi^2 2^{2h} e^{-i\pi h} \Gamma(2h) e^{2ih \psi}  (-q)^{-2h} = 4\pi^2 2^{2h} e^{i\pi h} \Gamma(2h)  \bar{w}^{-2h} .
\ee
The answer is well-defined for $h \in \frac{1}{2}\mathbb{Z}_+$ and diverges for $h \in \frac{1}{2}\mathbb{Z}_-.$ The latter divergences follow from the normalization of the conformal primary wave functions \eqref{ml} and were related to conformally soft poles in \cite{Pate:2019mfs,Puhm:2019zbl, Adamo:2019ipt}.

\bibliographystyle{plain}

\end{document}